\documentclass[prl,superscriptaddress,showpacs,showkeys,preprintnumbers,amsmath,amssymb,12pt]{revtex4}

\usepackage{dcolumn,amsmath,xspace}

\usepackage{graphicx}
\usepackage{dcolumn}
\usepackage{amsmath}

\usepackage{units}

\newcommand{\sixrt}{\ensuremath{(6\sqrt{3}\!\times\!6\sqrt{3})\text{R}30}~}

\begin{document}
\title{Growth mechanism for epitaxial graphene on vicinal 6H-SiC(0001) surfaces\\}

\author{M. Hupalo}
\affiliation{Department of Physics and Astronomy Iowa Sate University Ames Laboratory USDOE Ames IA 50011\\}
\author{E. Conrad}
\affiliation{The Georgia Institute of Technology, Atlanta, Georgia 30332-0430, USA\\}
\author{M. C. Tringides}
\affiliation{Department of Physics and Astronomy Iowa Sate University Ames Laboratory USDOE Ames IA 50011\\}

\begin{abstract}
The inability to grow large well ordered graphene with a specific number of layers on SiC(0001) is well known. The growth involves several competing processes (Si desorption, carbon diffusion, island nucleation etc.), and because of the high temperatures, it has not been possible to identify the growth mechanism. Using Scanning tunneling microscopy and an initially vicinal 6H-SiC(0001) sample, we determine the sequence of microscopic processes that result in the formation of bilayer graphene. Adjacent steps retract with different speeds and the carbon released produces characteristic "fingers" when a step of higher speed catches up with a slower moving step. These processes are based on different rates of Si desorption from steps and can be avoided if faster heating rates are used. We also show that faster rates lead to single layer graphene films extending over many microns.
\end{abstract}
\vspace*{4ex}

\maketitle
\newpage

The understanding of the structural and electronic properties of epitaxial graphene grown on SiC has proceeded rapidly since the suggestion that this material is a viable candidate for post CMOS electronics.\cite{Berger04,Berger06,Hass_JPhyCM_08} Graphene grown on either of the two polar faces of hexagonal SiC, the SiC(0001) silicon terminated surface (Si-face) and the SiC$(000\bar{1})$ carbon terminated surface (C-face), shows that these films behave like isolated graphene sheets.\cite{Berger06,Sadowski_PRL_06,Wu07,de Heer_SSC_07} While Si-face graphene films are more easily grown in UHV and have therefore been more extensively studied, the quality of these films has never achieved device levels.\cite{Hass_JPhyCM_08} Typically graphene grown in UHV on the Si-face is hampered by a high degree of SiC substrate roughening that leads to SiC terraces that are less than 50nm across.\cite{Hass_JPhyCM_08,Seyller_SS_06,Riedl_PRB_07,Hass_APL_06} On the other hand these films have a number of important advantages over C-face films. First, Si-face graphene is epitaxial with \sixrt periodicity as observed by low energy electron diffraction (LEED).\cite{vanBommel75,Forbeaux_PRB_98} The second advantage is that graphene growth in UHV on the Si-face of SiC is relatively slow compared to the C-face \cite{Hass_JPhyCM_08} and tends to be approximately 1-5 layers thick. However, to make devices from Si-face graphene requires a substantial improvement in its growth in order to produce graphene layers over large areas.

In this paper, we present STM data showing the kinetic processes that lead to bilayer ($G2$) and single layer ($G1$) graphene growth in UHV on the Si-terminated 6H-SiC(0001). We show through a series of experiments where the sample is heated in several steps (each step corresponding to an annealing time of ~30 sec at 1200C) that adjacent single steps have different evaporation rates. In this process $G2$ films form that extend over SiC terraces that are least 150nm wide. The increase of the terrace by factor of 3 (with the graphitization confined in only two layers) presents a vast optimization when both the lateral and vertical quality of the layer is considered. Recently, similar improvements have been found in high pressure furnace growth of Si-face films.\cite{Emtsev_ConMat_08} The different step evaporation rates combined with the observed asymmetric carbon diffusion parallel versus perpendicular to step edges leads to graphene "fingers" emanating from the step edges. To avoid the formation of $G2$ layers, faster one-step heating rates were used that result in large $G1$ domains. These experiments reveal that an initial surface containing only single SiC bilayer steps is a great advantage since they allow finer control of the Si evaporation rate and therefore finer control of graphene nucleation and growth.

The substrates used in these experiments were 6H-SiC(0001) purchased from Cree, Inc.\cite{Cree}. The samples were graphitized in UHV ($P\!\sim\! 1\times10^{-10}$torr) by direct current heating of the sample to $\sim\!1200C$ measured with an optical pyrometer. As seen in STM images the initial ungraphitized SiC substrate has a regular series of SiC bilayer steps (0.25nm high) with an average terrace length $w = 50$nm. Note that $\text{H}_2$ etching (the common pre-cleaning method used in epitaxial graphene growth \cite{Hass_JPhyCM_08}) leaves a surface with predominately 3-bilayer and 6-bilayer steps.\cite{Hass_JPhyCM_08} As we'll show below, it is the starting surface of single bilayer steps that allows the kinetics of the graphene growth to be observed in these studies.

\begin{figure}
\begin{center}
\includegraphics[width=8.0cm,clip]{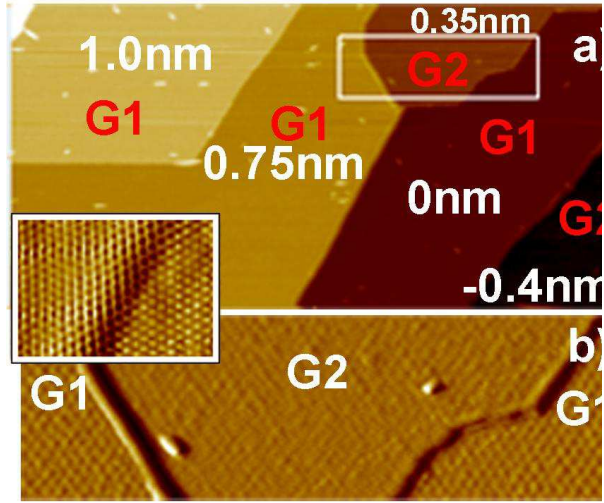}
\end{center}
\caption{The upper image is a $250\times 125 \text{nm}^2$ area STM image showing different heights on a graphitized 6H-SiC(0001) sample. The step heights can be written in terms of $s = 0.25$nm or $g = 0.35$nm and can be used to identify $G1$ or $G2$ graphene coverages. The lower image shows the contrast change in the \sixrt reconstruction between a $G1$ and $G2$ layer. The small inset to the left shows that graphene grows over a step.} \label{F:define_steps}
\end{figure}

In order to follow the graphitization process regions of different graphene thickness must be identified. This is done using both contrast changes in the \sixrt modulation intensity associated with single and double graphene layers,\cite{Riedl_PRB_07} and with relative step heights changes between different regions. This is demonstrated in Fig.~\ref{F:define_steps} where a graphitized sample shows a number of different surface heights. The bilayer imaged in the lower panel of Fig.~\ref{F:define_steps} shows a reduced modulation in $G2$ compared to $G1$ as previously demonstrated by Reidl et al.\cite{Riedl_PRB_07} All the step heights marked in Fig.~\ref{F:define_steps} (0.25nm, 0.75nm, 0.4nm 0.33nm) can be explained in terms of two heights: a bilayer SiC step $s = 0.25$nm, and a graphene step $g = 0.33$nm. For instance the 0.75nm step is $3s$ and the 0.4nm step is $3s-g$. It is important to note that the $G2$ layer forms between the SiC interface and $G1$. This is demonstrated in the small inset in Fig.~\ref{F:define_steps} were the top graphene layer ($G1$) is shown growing uninterrupted over a graphene step. This effect has also been observed by Lauffer et al.\cite{Lauffer_PRB_08}

\begin{figure}[htbp]
\begin{center}
\includegraphics[width=4.0cm,clip]{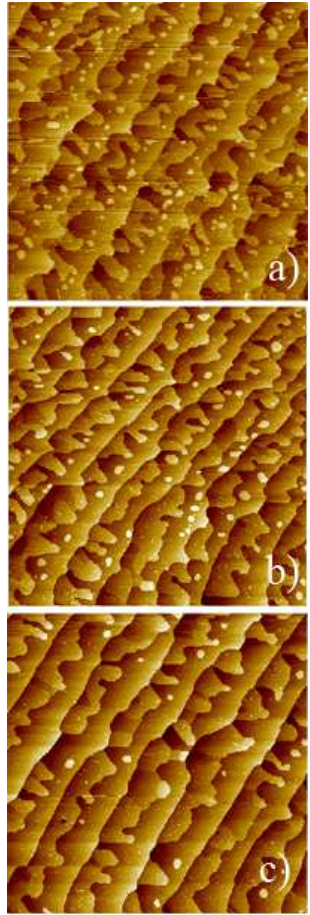}
\end{center}
\caption{Three STM images of $G2$ graphene growth after short 30sec heating steps to 1400C. The images show the transition from single to triple SiC bilayer steps. Three adjacent steps move with different speeds to form (a) $G2$ islands followed by (b) $G2$ "fingers" and finally (c) a continuous $G2$ layer. The size of the images is $1\mu\text{m}\!\times\!1\mu \text{m}$.} \label{F:fingers}
\end{figure}

The growth of the $G2$ layer is particularly unusual on SiC because stoichiometry requires the carbon contained in 3.14 SiC bilayers [$(2/a_\text{G}^2)/(1/a_\text{SiC}^2)= 3.139$] to form a single graphene sheet. This requirement makes layer-by-layer growth complicated. Indeed, we find that the formation of $G2$ is accompanied by a transition from single SiC bilayer steps to double bilayer steps and ultimately to triple SiC bilayer steps with terraces approximately three times the starting terrace length $w$, i.e. $3w \sim 150$nm. The change in terrace length preserves the starting vicinality of the surface. A detailed look at this transition reveals a number of important characteristics of $G2$ growth. The three growth stages are shown in STM images in Figs.~\ref{F:fingers}(a)-(c). During the first stage, graphene islands form. Shortly after this phase begins a partial transformation of some single steps to double steps occur. This second stage is accompanied by graphene "fingers" that appear growing away from the SiC steps edges. In the last stage a transition to triple steps is complete and the island and "fingers" have merged.

The details of Si desorption from the steps can be understood be correlating the $G2$ area formed with the local changes in the SiC substrate vicinality. During the formation of the $G2$ film in Fig.~\ref{F:fingers}, the SiC surface changes from single to double to triple bilayer steps. The step height changes indicate that different types of single SiC bilayer steps release carbon at different rates causing them to retract at different speeds. A fast evaporating single step will catch up to a slower single step to form a double bilayer step. The double step subsequently merges with the slowest third step to form a triple bilayer step. At this point there is enough carbon released in the retracting triple step to forms a continuous $G2$ layer. In contrast, if all steps retracted by evaporating Si at the same rate, there would be a constant concentration of carbon atoms that would lead to a steady rate of $G2$ nucleation events occurring randomly throughout all terraces (i.e. the first stage of island nucleation in Fig.~\ref{F:fingers}(a) would continue indefinitely). This scenario would produce a rough graphene layer contrary to the surface observed in Fig.~\ref{F:fingers}(c).

Figure~\ref{F:fingers}(a) shows that in the first stage of growth islands primarily nucleated within the first half of the terrace closest to the single step edge (in a length $w/2 \sim 25$nm). The islands in Fig.~\ref{F:fingers}(a) tend to be parallel with the step edges. This indicates that carbon diffusion is faster along steps rather than perpendicular to steps. More detailed statistics on island size and spacing find that the average island diameter, $<\!D\!>$, is 21.5nm while the average spacing, $<\!L\!>$, is 72nm. This implies that the average coverage in the region $w/2$ from the single step edge is $\theta\!=<\!D\!>\!/\!<\!L\!>=\!0.33$ML. This is in excellent agreement with the expected 1/3ML coverage expected because it takes the carbon released $\sim 3$ SiC bilayers to produce a single graphene layer.

\begin{figure}[htbp]
\begin{center}
\includegraphics[width=8.0cm,clip]{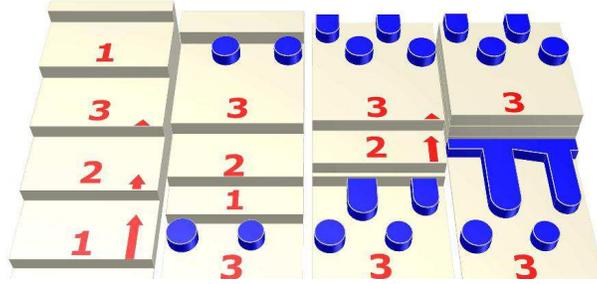}
\end{center}
\caption{A growth model showing schematically the processes observed in Fig.~\ref{F:fingers}. Three adjacent steps having different retraction speeds (indicated by the arrow length) generate successively $G2$ islands "fingers and a continuous $G2$ layer.} \label{F:Step_Schem}
\end{figure}

The growth sequence is shown schematically in Fig.~\ref{F:Step_Schem}. Three adjacent steps evaporate Si and release carbon as they are retracting. Based on the results of Fig.~\ref{F:fingers} we infer that the retracting speed of step 1 is larger than the retracting speed of 2, which in turn is larger than the retracting speed of step 3. The initial $G2$ islands begin to nucleate in the area exposed during the retraction of the first half of the terrace associated with the fast moving step 1. Because step 1 retracts faster it merges with step 2 to form a double SiC bilayer step. Before merging, the increased carbon released from the second half of the terrace associated with step 1 and the carbon released from step 2 combine to form "fingers". Eventually the last slow bilayer step 3 catches up to the retracting double step to merge into a triple bilayer step.

The process outlined in Fig.~\ref{F:Step_Schem} makes specific predictions about the areas covered by $G2$ at various stages of growth. These predictions can be checked against STM images. The "finger" coverage, when the triple step conversion is completed, is due to carbon released from two sources. The first carbon source is the conversion of the second half of the fast step 1 terrace width ($w/2$) after islands have formed (see the third panel of Fig.~\ref{F:Step_Schem}). The second carbon source come from the dissolving 2-bilayer terrace associated with step 2 of Fig.~\ref{F:Step_Schem} (i.e, the carbon in $2w$). Because a dissolving bilayer terrace produces 1/3 of a graphene layer, the "finger" coverage in a terrace $3w$ long should be $(1/3)(2w+w/2)/3w = 5/18$ML = 0.28. The measured "finger" coverage from Fig.~\ref{F:Step_Schem}(a) is 0.37ML. The final stage (when the triple bilayer step retracts to the edge of the initial islands) requires the carbon in $3w+2w+1w$ bilayers to be converted into $G2$ with an area proportional to $(1/3)(3w+2w+1w)$. This amount is spread over an area proportional to $3w$, so that the $G2$ coverage should be $\theta = (1/3)(3w+2w+1w)/3w = 2/3$ML. The experimental value in Fig.~\ref{F:fingers}(c) is found to be $\theta = 0.62$ML. Note that the initial finger diameter and finger separation are the same as the initial island diameter and island separation, $D = 20$nm and $S = 72$nm respectively. The finger separation remains unchanged as more carbon is released, but the "finger" width increases to $D = 40$nm. The graphene area produced along a step of length $S$ between the time the "fingers" begin to form and the time the triple step transition starts is $A = 1/3(w/2+2w)S = 3\times10^3\text{nm}^2$. Using this area, we can estimate the $G2$ "finger" length to be $L = A/D = 75$nm. This value is in good agreement with the 87nm length measured in Fig.~\ref{F:fingers}(b).

These experiments raise a number of interesting questions about the graphene growth mechanism. It is known that before graphitization a high carbon density \sixrt surface forms and remains only slightly perturbed after graphitization.\cite{Riedl_PRB_07,Johansson_PRB_96,Hass_CondMat_08} X-ray measurements indicate that this layer contains $\sim 2/3$ML of the carbon.\cite{Hass_JPhyCM_08,Hass_CondMat_08} It is entirely possible that breaking the Si-C bonds in one or two additional bilayers and the subsequent diffusion of Si through the \sixrt interface can release the additional carbon needed to complete $G1$. The additional kinetic barrier of Si diffusion through the first graphene layer prohibits the easy formation of $G2$ and requires Si to be released from steps as found in the current experiments. It should be remembered that the kinetics of diffusing Si through this first graphene layer is slower than Si diffusion through bulk graphite because this first graphene layer is more tightly bound to the interface.\cite{Hass_CondMat_08,Varchon_PRL_07,Mattausch_PRL_07}

\begin{figure}[htbp]
\begin{center}
\includegraphics[width=8.0cm,clip]{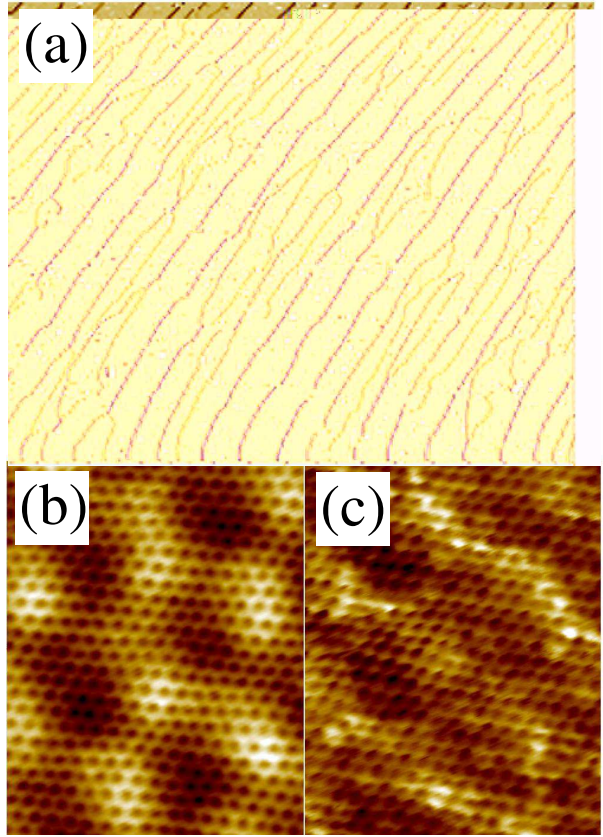}
\end{center}
\caption{(a) $2.5\times 2.0\mu \text{m}^2$ STM image where $85\%$ of the surface is a single ($G1$) graphene film. Because the annealing rate is faster than in Fig.~\ref{F:fingers}, very little $G2$ graphene is formed and no islands or "fingers" are observed. (b) and (c) $4.33\times 5.5\text{nm}^2$ image at different tunneling conditions -0.75V 0.1nA (b) and 0.75V 1nA (c) showing the dramatic dependence of the reconstruction modulation that confirms the film is a $G1$ layer.} \label{F:G1_Good}
\end{figure}

The series of annealing cycles at low temperatures used in the experiments discussed above have identified the importance of step retraction to control Si desorption at steps. On the other hand, heating rapidly to the growth temperature, changes the kinetics in a way that causes $G1$ to nucleate faster. The surface produced by this rapid heating is shown in Fig.~\ref{F:G1_Good}. Figure~\ref{F:G1_Good}(a) shows an STM image of a graphitized surface containing primarily ($85\%)$ a $G1$ film. Figures~\ref{F:G1_Good}(b) and (c) shows a closer look at the \sixrt reconstruction in this film at two different tunneling voltages -0.75V and 0.75V, respectively. The film is identified as a monolayer graphene sheet because of the strong dependence of the \sixrt  modulation with the bias voltage. That is, Fig.~\ref{F:G1_Good}(c) is more disordered than Fig.~\ref{F:G1_Good}(b); a result previously observed by Mallet et al.\cite{Mallet_PRB_07} Note the high degree of substrate order and graphene film thickness uniformity over areas larger than $2.5\mu\text{m}\times 2\mu\text{m}$ [see Fig.~\ref{F:G1_Good}(a)]. While the SiC terrace sizes are $\sim 100$nm (limited only by the original sample miscut) in Fig.~\ref{F:G1_Good}(a), the actual graphene order is significantly bigger because graphene grows over the SiC steps\cite{Seyller_SS_06}. Because the growth of $G1$ is so fast, the intermediate processes leading to the formation of this highly ordered first graphene layer (i.e. how Si is released) have not been identified.

Graphene preparation is still a hotly debated problem with key controlling barriers yet to be identified. We elucidate for the first time several kinetic processes that play a play a crucial role in the formation of the first and second graphene layers on the Si-face of SiC; Si desorption through steps and the fine balance of different step evaporation rates and the diffusion anisotropy of carbon. We show that graphene domains can be grown that are an order of magnitude larger than those prepared by previous UHV preparation methods. The most important conclusion of the current experiments is that single steps are the controlling factors for Si desorption and that different SiC steps have different evaporation rates. This result is crucial to suppress random and uncorrelated nucleation events and to confine the growth to only two layers $G1$ and $G2$. This result explains why SiC surfaces prepared by $\text{H}_2$ etching leads to more disordered film. $\text{H}_2$ etching produces a starting substrate with multiple steps.\cite{Hass_JPhyCM_08} During graphitization, these receding multiple bilayer steps cause uncontrollable random graphene nucleation events resulting in low quality graphene layers. Although the origin of the different retraction speeds is not clear, similar differences between single step growth rates have been observed during CVD growth of SiC.\cite{Kimoto_JAP_97} TEM studies show step bunching from single to triple steps during SiC growth at high temperature ($\sim 1500C$). They find that the fast moving step has a lower surface energy (6meV/atom) than the two slower steps that merge into the triple step bunch.\cite{Kimoto_JAP_97} This suggests a similar scenario of step retraction that is outlined in Fig.~\ref{F:Step_Schem}.

\begin{acknowledgments}
We wish to thank A. Zangwill and D. Vvedensky for helpful discussions and a critically reading this manuscript. Work at the Ames Laboratory was supported by the Department of Energy-Basic
Sciences under Contract DE-AC02-07CH11358. Work at Georgia Tech was supported by a grant from the W.M. Keck Foundation and the National Science Foundation under Grant Nos. 0404084 and 0521041.
\end{acknowledgments}

\end{document}